\documentclass[twocolumn,amsmath,amssymb,10pt,nofootinbib,superscriptaddress]{revtex4}
\usepackage{ graphicx, float,amsmath, amsmath, amssymb}
\usepackage{xcolor} 
\usepackage[export]{adjustbox}
\usepackage[breaklinks, colorlinks, citecolor=blue]{hyperref}
\usepackage[labelsep=period]{caption}
\usepackage[normalem]{ulem}
\usepackage{mathtools}
\usepackage{siunitx}
\usepackage{soul}
\usepackage{physics}
\usepackage{minitoc}

\usepackage{bm}
\usepackage{xcolor}

\definecolor{alizarin}{rgb}{0.82, 0.1, 0.26}

\def\be{\begin{equation}}
\def\ee{\end{equation}}
\def\bea{\begin{eqnarray}}
\def\eea{\end{eqnarray}}
\def\bse{\begin{subequations}}
\def\ese{\end{subequations}}

\begin{document}
\setlength{\parindent}{0cm}

\title{Closer look at cosmological consequences of interacting group field theory}

\author{Maxime De Sousa}%
\affiliation{%
Laboratoire de Physique Subatomique et de Cosmologie, Universit\'e Grenoble-Alpes, CNRS/IN2P3\\
53, avenue des Martyrs, 38026 Grenoble cedex, France
}

\author{Aur\'{e}lien Barrau}%
\affiliation{%
Laboratoire de Physique Subatomique et de Cosmologie, Universit\'e Grenoble-Alpes, CNRS/IN2P3\\
53, avenue des Martyrs, 38026 Grenoble cedex, France
}

\author{Killian Martineau}%
\affiliation{%
Laboratoire de Physique Subatomique et de Cosmologie, Universit\'e Grenoble-Alpes, CNRS/IN2P3\\
53, avenue des Martyrs, 38026 Grenoble cedex, France
}



\date{\today}
\begin{abstract} 
Group field theory has shown to be a promising framework to derive cosmological predictions from full quantum gravity. In this brief note, we revisit the background dynamics when interaction terms are taken into account and conclude that, although the bounce is clearly robust, providing a geometrical explanation for inflation seems to be very difficult. We consider possible improvements and modifications of the original scenario and derive several limits on the parameters of the model, trying to exhaustively investigate the relevant features and weaknesses of the approach.
\end{abstract}
\maketitle

\section{Introduction}

Formally, group field theories (GFTs) are a class of tensor quantum field theories  generalizing matrix models to higher dimensions. In practice, GFT is a general formalism suitable for quantum gravity (see, {\it e.g.}, \cite{Oriti:2005tx,Oriti:2006se,Oriti:2009wn,Baratin:2011aa,Carrozza:2020akv} for introductions and reviews). It can be understood as a kind of second quantization of loop quantum gravity (LQG) \cite{Oriti:2013aqa,Oriti:2014uga}. General introductions to LQG can be found in \cite{Rovelli:2004tv,Rovelli:2014ssa,lqg3,Ashtekar:2021kfp,Rovelli:2022xsj}. Basically, excitations correspond to spin-network nodes (or to simplicial building blocks) that can be combined to obtain the usual spin-network states.\\

The framework has been successfully applied to cosmology \cite{Calcagni:2012vb,Gielen:2013kla,Calcagni:2014tga,Gielen:2016dss,Pithis:2019tvp,Marchetti:2020umh}. Among others, a remarkable result (obviously relying on quite heavy assumptions) is that the main consequence of loop quantum cosmology (LQC) (See, {\it e.g.}, \cite{lqc9}) -- that is a bounce instead of the big bang -- is recovered. The usual Friedmann equation is corrected by a density squared term with the opposite sign, which prevents the collapse into a singularity. The agreement is not perfect, which raises both technical and conceptual questions, but, basically, GFT and LQC converge to bouncing scenarios. \\

Recently, exciting results were derived on cosmological perturbations in GFT \cite{Gielen:2015kua,Marchetti:2021gcv}. Computing the primordial power spectra, which can be compared with cosmological microwave background (CMB) measurements, remains a major challenge. In this work, we focus on a less ambitious, but mandatory, point: the background dynamics when allowing interactions between the GFT blocks in the separate universe approach \cite{Gerhardt:2018byq}. This is exactly what has been investigated in \cite{deCesare:2016rsf}. It is an important step before considering seriously perturbations.\\

This note is devoted to a critical check of the remarkable results obtained in \cite{deCesare:2016rsf}. Beyond a mere critical reading of this pioneering work, we provide several suggestions for fruitful completions and underline their possible drawbacks. Section \ref{section:InteractingGFTCosmology} is devoted to a brief explanation of the model and of the main conclusion of \cite{deCesare:2016rsf}, that is the possibility to have a long enough stage of primordial inflation from a purely quantum geometrical origin. We then point out in Section \ref{section:DifficultiesGFTInflation} that, for different reasons, this result might, unfortunately, not hold. We generalize this statement in Section \ref{section:PossibleCure} and suggest a possible completion which, however, turns out to be inconsistent with the specific GFT requirements. We finally conclude in section \ref{section:latetime} with generic limits that might be derived in this framework and underline the main cosmological features of interacting GFT.

\section{Interacting GFT and cosmology}
\label{section:InteractingGFTCosmology}

In the group field theory condensate cosmology program, space-time is an emergent concept \cite{Oriti:2009zz}. We do not go here through the numerous subtleties of this very appealing picture and we simply summarize the results of \cite{deCesare:2016rsf} in a ``cosmologist-friendly" way without introducing the intermediate structures. In particular, we choose not to introduce the different fields 
appearing in the reasoning and to express all the relevant quantities as a function of the more intuitive GFT volume element, focusing on effective Friedmann-like equations. We ignore all non-mandatory technicalities, in particular the question of deparametrization. We occasionally refer to a massless scalar field $\phi$ which plays the role of a relational time. Details about the framework can be found in \cite{Oriti:2016ueo}.\\ 

The Hubble parameter is defined as:
\begin{equation}
    H \equiv \frac{1}{3}\frac{\partial_{\phi}V}{V^2}\pi_{\phi},
\end{equation}
where $\pi_{\phi}$ is the momentum associated with the scalar field assuming to fill the Universe, and $V$ is the expectation value of the volume operator, which is diagonal in the basis of spin representations \cite{Oriti:2015qva}. The effective Friedmann equation associated with the free GFT theory is well known \cite{Gielen:2013kla}. It basically reads, for a given $j$ representation of SU(2):
\begin{equation}
    H^2_{\text{free}} =  \kappa
 \biggl( \epsilon_m \, V^{-2} + \epsilon_E \, V^{-3} + \epsilon_Q \, V^{-4} \biggr) ,
 \label{Friedmann Free GFT}
\end{equation}

with $\kappa=8\hbar^2 Q^2/9$. The $\epsilon_i$-coefficients are given by:
\begin{equation}
    \epsilon_m = m^2/2, \quad
    \epsilon_E = V_j E \quad \text{and} \quad
    \epsilon_Q = -V_j^2 Q^2/2.
\end{equation}
The $Q$ parameter is the conserved charge associated to the phase of the GFT condensate wavefunction, so that $\pi_{\phi}=\hbar Q$, $m$ is a ratio of parameters entering the equation of motion of the field corresponding to the selected $j$ representation (see, once again, \cite{deCesare:2016rsf} for details), and $E$ is the so-called GFT energy \cite{deCesare:2016axk}. At the effective level of this study they can be considered as fundamental parameters. The volume $V_j$ is given by $V_j = V_{Pl}j^{3/2}$ which will be approximated, in the following, as $V_j\sim V_{Pl}$. It is well known that Eq. (\ref{Friedmann Free GFT}) leads to a bounce as a replacement of the initial singularity \cite{Oriti:2016qtz, Oriti:2016ueo}. It can be immediately be seen in the opposite signs of the $V^{-2}$ (or $V^{-3}$) and $V^{-4}$ terms: at some point the sum vanishes. Both from a purely heuristic viewpoint or from the specific GFT relations, it is reasonable to assume that $V \sim V_{Pl}$ at the bounce. Since at this moment $H = 0$, one is left with the following relation between the $\epsilon_i$ coefficients: $\epsilon_m + \epsilon_E V_{Pl}^{-1} + \epsilon_Q V_{Pl}^{-2} = 0$.
It should however be kept in mind that usual GFT condensate equations might not be valid beyond the so-called \textit{mesoscopic regime} \cite{Oriti:2016qtz} and great care should be taken in their usage.

Taking into account interactions between the fundamental building blocks is an important refinement of the initial scenario to understand the dynamics of the resulting spacetime. As shown in \cite{deCesare:2016rsf}, two leading interaction terms are expected to appear, with couplings $\lambda$ and $\mu$, in the potential entering the  equation of motion of the modulus of the $\sigma-$field corresponding to the particular spin$-j$ representation chosen. This translates into an effective Friedmann equation

\begin{equation}
    H^2 = H^2_{\text{free}} + \kappa \biggl( \epsilon_\lambda \, V^{n/2 - 3} + \epsilon_\mu \, V^{n'/2 - 3}  \biggr),
    \label{FriedMarco}
\end{equation}
where the $\epsilon$-coefficients are defined as
\begin{equation}
    \epsilon_\lambda = -\frac{\lambda}{n} V_j^{1-n/2} \quad \text{and} \quad \epsilon_\mu = -\frac{\mu}{n'} V_j^{1-n'/2}.
\end{equation}
$n$ and $n'$ being integers related with the underlying nature of the interactions. 

The Raychaudhuri-like equation reads

\begin{widetext}
    \begin{equation}
\frac{\ddot{a}}{a} =  -\frac{\kappa}{2} \left( 4 \epsilon_m  V^{-2} + 7\epsilon_E  V^{-3} + 10\epsilon_Q  V^{-4} + \alpha V_j^{1-n/2} V^{n/2-3}+ \beta V_j^{1-n'/2} V^{n'/2-3} \right),
\label{RaychauEq}
\end{equation}
\end{widetext}

with
\begin{equation}
    \alpha \equiv \biggl( 3 - 14/n \biggr) \lambda,
\end{equation}
and
\begin{equation}
    \beta \equiv \biggl( 3 - 14/n' \biggr) \mu.
\end{equation}

The study carried out in \cite{deCesare:2016rsf} shows that with $\lambda<0$, $\mu>0$, $n=5$, $n'=6$, and a strong hierarchy between $\mu$ and $|\lambda|$, one can get enough e-folds of purely geometrical inflation. The free case also leads to a phase of super-inflation (as in LQC, or more generally as in any bouncing scenario) but is always very brief and cannot account for a sufficient inflationary stage (as in LQC). \\

The usual requirement for the number or inflationary e-folds, $N>60$, which is taken as a baseline in \cite{deCesare:2016rsf}, has many well known motivations. If, however, one focuses on the most stringent one, that is the generation of the correct primordial power spectrum, it is worth underlying that the situation is intricate. The standard process for the generation of perturbations relies on the quantum fluctuations of the inflaton field. Although some aspects of the usual scenario can be understood in the framework of  effective theories of inflation \cite{Cheung:2007st}, somehow disentangling the question of perturbations themselves from the question of the process which generates them, it should be kept in mind that the derivation of the primordial power spectrum -- which is the key result of inflation -- is not anymore straightforward. Otherwise stated: it is not clear that it is either necessary or sufficient to have 60 e-folds of "inflatonless inflation".

In the following, to make comparisons easier, we use the same specific definition for the acceleration as in \cite{deCesare:2016axk}:
\begin{equation}
\mathfrak{a} \equiv \frac{\partial^2_{\phi}V}{V}-\frac{5}{3}\left( \frac{\partial{\phi}V}{V} \right)^2.
\end{equation}
Explicitly, it can straightforwardly be shown to be given by:
\begin{equation}
\mathfrak{a}=
-2\left( \frac{V}{V_j}\right)^{-2}\left( P+\alpha \left( \frac{V}{V_j} \right)^{n/2+1} + \beta \left( \frac{V}{V_j} \right)^{n'/2+1} \right),
\label{acc}
\end{equation}
where $P=4m^2\left( V/V_j \right)^2+ 14 E 
\left( V/V_j \right) - 10 Q^2$ corresponds to the free case. The sign of $\mathfrak{a}$ is the same as the sign of $\ddot{a}/a$ ($a$ being the scale factor), however the actual value is not the same. The relation reads:
\begin{equation}
\frac{\ddot{a}}{a}=\frac{1}{3}\left( \frac{\pi_{\phi}}{V} \right)^2 \mathfrak{a}.
\label{aPointPointFrak}
\end{equation}

\section{Difficulties with GFT inflation}
\label{section:DifficultiesGFTInflation}

Assuming that interaction terms dominate during most of the dynamics, the number of inflationary e-folds is given by
\begin{equation}
    N = \frac{1}{3} \ln \biggl( \frac{V_\text{end,inf}}{V_B} \biggr),
\end{equation}
where $V_\text{end,inf}$ and $V_B$ are, respectively, the volume at the end of inflation and the volume at the bounce. The former can easily be computed from Eq.\eqref{aPointPointFrak} requiring $\ddot{a}=0$. This leads to 

\begin{equation}
     N =  \frac23 \left( n' - n \right)^{-1} \biggl[ \ln \left( - \frac{\, \, 3-14/n'}{3 - 14/n} \right) - \ln \left( \lambda \mu\right) \biggr].
\end{equation}

Obviously, a high level of fine-tuning of the GFT free parameters is required in order to generate as much inflation as needed. Remarkably, this is however possible. In this model, the $n-$interaction term triggers inflation and the $n'-$interaction term ends inflation. One should notice that some of the usual intuitions about inflation do not hold in this context. For example, in the standard case, a positive acceleration requires terms in the Friedmann equation scaling as $a^{-\alpha}$ with $\alpha<2$. Although the specific case $n=5$ fulfils this condition, it is not anymore required here, as we shall see later on.\\

Importantly, in the GFT inflationary scenario, the Universe undergoes a recontraction after inflation. This can easily be seen: the requirement of having a sign change of the acceleration (so as to end inflation), from positive to negative, implies that the $n'$ term, which inevitably dominates for large volumes (as $n'>n$), must have a negative coefficient in Eq. (\ref{acc}), that is $\beta>0$. This is in agreement with values selected in \cite{deCesare:2016rsf}. However, considering now the Friedmann equation (\ref{FriedMarco}), it is clear that the $n'$ term will also dominate for large volumes. As it appears with a negative coefficient ($\mu>0 \Rightarrow \epsilon_\mu<0$), this will trigger a contraction of the Universe.

It is possible to calculate the ratio between the scale factor at recontraction, that is at the end of expansion, and the scale factor at the end of inflation, that is at the end of acceleration, $r = a_{\text{end,exp}}/{a_\text{end,inf}}$. It reads:

\begin{equation}
     r = \biggl( 3 \frac{\, n'/2 - 1}{\,n/2 - 1} \biggr)^{2(n'-n)/3}.
\end{equation}
The case considered in 
\cite{deCesare:2016axk} leads to $r \approx 5/2$. Any other reasonable choice for the parameters also lead to $r\sim 1$. This means that space starts to contract very soon after the end of the inflation and our Universe does not have time to exist. Otherwise stated: the $n'-$interaction term has to dominate in the expression of the acceleration (say in the Raychauduri equation) to end inflation but when it does so, it inevitably also dominates soon after in the expression of the expansion speed (say in the Friedmann equation), leading to an unavoidable fast contraction of the Universe. The model does not lead to anything looking like our world.\\

Let us underline another problem with this approach. The scenario advocated in \cite{deCesare:2016rsf} requires $n\ge 5$ and $n'>n$. When the $n'-$interaction term dominates in the GFT Friedmann equation, describing the post-inflationary universe, one is therefore led to
\begin{equation}
    H^2\propto \frac{1}{a^\alpha},
\end{equation}
with $\alpha<3/2$. This means that radiation -- with an energy density scaling as $a^{-4}$ in the Friedmann equation -- will never naturally overcome the GFT interaction term. Even ignoring the recontraction phenomenon, this is an issue, especially when taking into account that no obvious reheating process is here expected. Of course, subtle mechanisms might allow for a transition between a GFT dominated stage and the usual radiation dominated one. The fact remains that the scaling laws do not favour a natural evolution from one to the other. \\

More generally, one might wonder if  
describing the interacting part of the potential entering the equation of motion of the system with a generic potential like
\begin{equation}
    U = \sum_{n>2} \frac{\lambda_n}{n} \left( \frac{V}{V_j}\right)^{n/2},
    \label{GeneralPotential}
\end{equation}
with $\lambda_n \in \mathbb{R}$, might solve the situation. This is actually not the case. Terms such that $\lambda_n ( 14/3-n ) \geq 0$ lead to a positive acceleration whereas others lead to a negative acceleration.
Either terms leading to a positive acceleration always dominate and inflation never ends, or a term leading to negative acceleration becomes, at some point, the leading one after inflation. In such a case -- this is the key-point -- the Universe inevitably begins to contract very soon. Notice that the corresponding term enters the Friedmann and Raychauduri equations with the same power of $V$. This process will not end before a ``small volume" bounce takes place again. There is no way out. Interestingly, one can notice that the acceleration (the real one, that is $\ddot{a}/a$) does tend toward a constant if the dominant term is associated with $n=6$, it tends toward $0^+$ if $n<6$, and it tends toward infinity if $n>6$.

\section{A possible completion}
\label{section:PossibleCure}

Let us now investigate whether the previously mentioned difficulties could be overcome without giving up a geometrical origin for inflation. This might require to loosen the constraints on the signs of the different parameters given in \cite{deCesare:2016rsf}. Using notations of \cite{deCesare:2016rsf} (we do not need to go here through a detailed explanation), let us mention that the requirement $\omega'>0$ is robust as it ensures that the potential entering the GFT condensate action is bounded from below \cite{Oriti:2016qtz,Oriti:2016ueo}. So is the requirement $A/B>0$ as it is grounded in the compatibility with the free case \cite{deCesare:2016axk}. However, the other constraints are basically related to phenomenological constraints that can, in principle, be relaxed. One can even argue that choosing signs so that the late-time behavior of the universe recontracts instead of being accelerating is, actually, quite questionable.\\

We shall first wonder if the model can, through a refinement of the previous scenario, explain inflation in a self-consistent way, using only geometrical features.

The effective energy density associated with the $n-$interaction term scales as $\rho_n \propto a^{-\gamma}$, with $\gamma=-3n/2+9$ (corresponding to an effective equation of state $w=2-n/2$). As we have shown that relying on the second interaction term to end inflation inevitably leads to a recontraction (and conflicts with a natural overcome of radiation), we instead investigate now if radiation could, in itself, trigger the end of the geometrical inflation. Inflation would take place thanks to GFT interactions and the usual content (in this case a relativist fluid) would then take over the dynamics. For this scenario to work, one needs $\gamma >4$. This ensures that the interaction term leading to inflation gets diluted faster than radiation, allowing the transition to the usual hot Big Bang. This translates into $n<3.3$.

On the other hand, the $n-$interaction term should, at some point, dominate over $P(V)$ in the acceleration (once again, we use the volume as a cosmologist-friendly parameter, instead of the field $\rho$ of \cite{deCesare:2016rsf}). 
As $P(V)$ contains a term scaling as $V^2$ in the acceleration whereas the $n-$interaction term behaves as $V^{n/2+1}$, this requires $n>2$. 
Together with the previous condition, this leads to $n=3$. 

Finally, to lead to a positive acceleration (by definition of inflation), one should ensure that  $\alpha <0$. As $\alpha=(3-14/n)\lambda\approx -1.7\lambda$, this demands $\lambda>0$. We deliberately choose to first ignore this constraint to suggest a general argument.\\

The second interaction term (the $n'-$one) is not anymore required to end inflation and the duration of the primordial accelerated expansion phase is now not fixed by the ratio between $\alpha$ and $\beta$. Inflation would last forever without the radiation term which, at some point, overcomes the $n-$interaction one.\\

In GR, the Friedmann equation relates the Hubble parameter to the energy density:
\begin{equation}
    H^2_{GR}=\frac{8\pi G}{3}\rho_X.
\end{equation}

The density $\rho_X$ is the physical parameter that dictates the expansion rate of the Universe. It can, of course, be expressed as a function of the scale factor, when taking into account the equation of state and  introducing a proper normalization. For example, using the $\propto a^{-3}$ dilution in the matter era and $\propto a^{-4}$ dilution in the radiation era, the Friedmann equation can be approximated by:
\begin{equation}
    H^2_{GR}=\frac{8\pi G}{3}\Omega_0\rho_c\biggl(1+z_{eq}\biggr)^3\biggl( \frac{a_{eq}}{a}\biggr)^4,
    \label{hgr}
\end{equation}
where $\rho_c$ is the critical density, $\Omega_0\approx0.3$, $z_{eq}$ and $a_{eq}$ are the redshift and scale factor at the equilibrium time.

In GFT, the situation is different. The physical parameter which dictates the expansion rate -- at least in a  phenomenological view -- is the volume. The Hubble parameter depends upon the volume. Which volume ? Obviously {\it neither} the volume of a fiducial space region, which is arbitrary and cannot fix the expansion speed {\it nor} the Hubble volume which is unrelated with quantum gravity effects. One should instead consider the volume of a fundamental ``patch". As previously stated, it is reasonable to assume that $V\sim V_{Pl}$ at the bounce and to write:
\begin{equation}
    V=V_{Pl}\left( \frac{a}{a_B} \right)^3,
\end{equation}
where $a_B$ is the scale factor at the bounce. 
The GFT Friedmann equation therefore reads, when the $n-$interaction term dominates:
\begin{equation}
    H^2_{GFT}=\frac{8\hbar^2Q^2}{9}\left( -\frac{\lambda}{3} \right) V_{Pl}^{-2}\left( \frac{a}{a_B} \right)^{-9/2},
    \label{hgft}
\end{equation}
where $a_B$ is the scale factor at the bounce. This reasoning is, as it should be, very different from what is performed in usual cosmology. Once again, as if there were no fundamental length scale in the problem, the equation is invariant by reparametrization of the scale factor. We have here assumed that the number of e-folds associated with the pre-inflationary dynamics is negligible, which is indeed the case. 

The transition between both regimes takes place when the Hubble rate given by Eq. (\ref{hgr}) matches the one given by Eq. (\ref{hgft}). This precisely corresponds to the end of inflation.\\

A rather subtle point should raised. In the initial scenario of \cite{deCesare:2016axk}, inflation ends due to a second GFT interaction term. This made the number of e-folds {\it a priori} calculable from the parameters of the model. In the version we propose here, inflation ends due to the usual radiation term. This, however, comes with an interesting feature. For both equations (\ref{hgr}) and (\ref{hgft}), the scale factor can be normalized arbitrarily. But their simultaneous use becomes tricky as one requires physical values to be set at the bounce whereas the other needs them them to be determined at the equilibrium time (or anywhere in the post-GFT phase). Otherwise stated: there is no way the GFT Hubble rate can be fixed using a scale factor normalised to a post-GFT phase. This brings an interesting freedom in the game. 

This important issue is {\it not} specific to GFT. An analogous situation was met in \cite{Alesci:2018qtm}. Matching the usual cosmological behaviour where the choice of a fiducial volume it totally arbitrary with a quantum gravity approach, where there is a fundamental length scale and where the absolute value of the considered volume does matter, is tricky. The way we proceed here is naive but fills the gap in a phenomenologically consistent way.\\

In practice, terms in Eq. (\ref{hgr}) are approximately known. Let us call $a_t$ the scale factor at the transition between the GFT (inflationary) phase and the GR (radiation dominated) phase. The ratio $(a_{eq}/a_t)$ can be better expressed as $(T_{RH}/T_{eq})$ where $T_{RH}$ is the temperature at the end of inflation and $T_{eq}$ is the equilibrium temperature. We shall fix it following the usual guess is $T_{RH}/T_{eq}\sim 10^{25}$, although a quite wide interval in actually allowed (especially toward lower values). The transition between the GR and the GFT regimes corresponds to:
\begin{equation}
    \lambda Q^2=-\left( \frac{a_t}{a_b}\right)^{9/2}\frac{9\pi G}{\hbar^2}V_{Pl}^2\Omega_0\rho_c\biggl(1+z_{eq}\biggr)^3\biggl( 
\frac{T_{RH}}{T_{eq}}\biggr)^4.
\label{Lambda Q^2}
\end{equation}
Once the desired number of inflationary e-folds, which is very close to $\text{ln}(a_t/a_B)$, is set, the value of $\lambda Q^2$ is fixed. This is directly related to the GFT condensate action and the effective potential describing the interactions.\\

This methodology is quite appealing. It is however {\it not} viable for at least two reasons.

The first one is the most obvious: $\lambda$ does not have the right sign. The GFT Friedmann equation, when the $n-$interaction term dominates, requires $\lambda > 0$. This contradicts the requirement of this analysis, as Eq. (\ref{Lambda Q^2}) obviously requires a negative value of $\lambda$. We, however, believe that it was worth going through this discussion as it might be helpful in another context, suggesting a possible ``graceful exit".

The second one is less trivial. In this model, the dynamics is governed by GFT before the transition time, when $a<a_t$. However radiation is still present to allow the end on inflation. By construction, the volume entering the GFT Friedmann equation is roughly planckian at the bounce. This is obviously consistent. As far as {\it distances} (to whatever power) are concerned, everything is fine. However, the radiation energy density does grow exponentially fast during inflation (when thinking backward in time). In the usual inflaton-based scenarios this does not happen as $\rho_{\text{inflaton}}\sim \text{constant}$ when $w_{\text{inflaton}} \sim -1$. Here, inflation is purely geometrical and radiation, which is still present, leads to a highly -- and probably problematic -- transplackian energy {\it density} at the bounce. This is not, in itself a GFT issue, this is a consequence of not relying on the usual reheating process.

Finally we should also point out the way usual matter is combined with the GFT dynamics is of course obviously very naive. This is a first try in this direction aimed at fixing ideas.

\section{Late time behavior and constraints}
\label{section:latetime}

Let us now consider the interacting GFT approach to cosmology without trying to explain inflation geometrically. The quantum geometrical ingredients would explain the bounce while inflation would be ensured by a massive (or more complicated) scalar field. Describing consistently the dynamics of this field is not yet fully clear in GFT. We therefore keep using the toy-model  suggested before. In this case, no interaction term is required in GFT and the free case is sufficient. Assuming, naturally, that the (free) term with the smallest inverse power of the volume soon dominates the GFT dynamics, one would be led to:
\begin{equation}
    \frac{8\hbar^2Q^2}{9}\frac{m^2}{2V_{infl}^2}\sim\frac{8\pi G}{3} \rho_{\varphi},
\end{equation}
where $\varphi$ is the inflaton field, $\rho_{\varphi}$ is its energy density during inflation, and $V_{infl}$ is the GFT volume at the transition time between the (free) GFT domination and inflation. Even if $\rho_{\varphi}$ is assumed to be known (which is an optimistic hypothesis), it is clear that the constrain only involves the combination $mQ/V_{infl}$, which is not very stringent. If we follow the previous logic, it is reasonable to assume that $V_{infl}$ is not much greater than $V_{Pl}$, which make the constraint more interesting.\\

As well known \cite{Oriti:2021rvm} and quite obvious, an interaction with $n=6$ leads to a term which does not depend upon the volume in the Friedmann equation, hence an effective cosmological constant. In this case, one can take advantage of the fact there is one parameter less entering the game and investigate if this could explain the contemporary acceleration of the Universe. To fix order of magnitude, one gets:
\begin{equation}
    \frac{8\hbar^2Q^2}{9}\left( \frac{-\lambda}{6} \right) V_{Pl}^{-3/2}\sim\frac{8\pi G}{3} \rho_c,
\end{equation}
where $\rho_c$ is the critical (or current) density. Obviously, the very small value of $\rho_c$ directly translates into a very small value for $\lambda$.  Although the current acceleration of the Universe can be explained this way, this does not help solving usual fine-tuning issues.\\

Finally, it could also be that there is a $n'-$interaction, leading to the recontraction of the Universe, but that this has just not yet taken place -- hence being in agreement with the real world. According to \cite{Andrei:2022rhi}, a recontraction could happen in less 0.3 Hubble time from now. This is very soon and it corresponds to $a_{rec}\sim 3.7a_0$. The corresponding limit reads
\begin{equation}
    \frac{8\hbar^2Q^2}{9}\biggl( \frac{\mu}{n'} \biggr) V_{Pl}^{1-n'/2} \left( V_{Pl} \left( \frac{3.7a_0}{a_B}\right)^3\right)^{n'/2-3}
    <\frac{8\pi G}{3} \rho_c.
\end{equation}
As the total number of e-folds between the bounce and now, $\text{ln}(a_0/a_B)$, is not known, this does not translate into a useful bound. It simply means that $\mu$ has to be tiny.

\section{Conclusion}

Interacting GFT opens exciting paths toward a full quantum gravitational understanding of the early Universe. Remarkably, the free theory cures the big bang singularity and provides a regular description of the ``small volume" regime. Interaction terms can play a role in the dynamics but we have shown that it seems difficult to use them to generate inflation. There are two independent reasons for this. First, the Universe inevitably undergoes recontraction soon after the end of inflation. Second, the scaling of the interaction term responsible for ending inflation hardly allows radiation to dominate at any time in the future.

We have suggested a kind of grateful exit, which could be used in other scenarios. The idea is to use interaction terms to trigger and maintain inflation but to rely on radiation to exit inflation. This scenario can however not be made compatible with GFT. Nevertheless, our conclusion is based on a naive approach and a proper treatment of radiation in GFT should be investigated in the future. This might cure the difficulties exhibited in this work.

Finally, some general limits on interaction parameters are derived so that the model remains compatible with observations.\\

Clearly, much remains to be done. The effective equations used throughout this work are not valid for all values of the underlying parameters. Improvements at this level might lead to substantial modifications of the claims made in this article.

\bibliography{refs.bib}

 \end{document}